\documentclass[12pt,preprint]{aastex}
\begin{document}

\title{Lack of observational evidence for quantum structure of
  space--time at Planck scales \footnotemark[1]}
\author{Roberto Ragazzoni}
\affil{INAF -- Astrophysical Observatory of Arcetri (Italy)\\
       Max Planck Institut fuer Astronomie -- Heidelberg (Germany)}
\email{ragazzoni@arcetri.astro.it}

\and

\author{Massimo Turatto}
\affil{INAF -- Astronomical Observatory of Padova (Italy)}
\email{turatto@pd.astro.it}

\and
\author{Wolfgang Gaessler}
\affil{Max Planck Institut fuer Astronomie -- Heidelberg (Germany)}
\email{gaessler@mpia.de}

\footnotetext[1]{Based on observations made with the NASA/ESA Hubble
Space Telescope, obtained from the data archive at the Space Telescope
Institute. STScI is operated by the association of Universities for
Research in Astronomy, Inc. under the NASA contract  NAS 5-26555.}

\begin{abstract}
It has been noted (Lieu \& Hillmann, 2002) that the cumulative affect of
Planck--scale phenomenology, or the structure of
space--time at extremely small scales, can be lead to the loss of phase
of radiation emitted at large distances from the observer.
We elaborate on such an approach and
demonstrate that such an effect would lead to an apparent blurring
of distant point--sources.
Evidence of the diffraction pattern from the HST observations
of SN 1994D and the unresolved appearance of a Hubble Deep Field
galaxy at z=5.34 lead us to put stringent limits on the effects of 
Planck--scale phenomenology.
\end{abstract}

\keywords{gravitation--time}

\section{Introduction}

It is generally believed that a description of the gravity consistent with quantum theory (quantum gravity) should imply properties of the space--time much different from the conventional ones, when the latter is being observed at the so--called {\it Planck scale}. Such a Planck scale is obtained as a combination of fundamental
constants and corresponds to
a characteristic length $l_P \approx 1.6\times 10^{-35}$ m and time interval $t_P\approx 5.4 \times 10^{-44}$ s given by:

\begin{equation}
l_P=ct_P = \sqrt{ \frac{G\hbar}{c^3} }
\end{equation} 

Space and time, when observed at such scales, are expected to exhibit a grainy, fuzzy or a foam-like structure, as depicted by several authors (see for instance Rovelli 1998, Garay 1998, Kempf 1999). The operational definition of measurement of a length or of a time--interval should be affected by such property of space--time (Wigner, 1957; Salecker \& Wigner. 1958; Adler \& Santiago, 1999) and one can conceive several {\it gedanken} experiments that should be affected by the so--called Planck Scale
Phenomenology (hereafter PSP, see Amelino--Camelia, 2001a).

In spite of the extremely small size of $l_P$, recently several authors pointed out that the {\it systematic accumulation} of such effect during the 
long journey of a photon propagating through
a space--time affected by PSP could
lead to observable consequences.
Several possible measurements has been proposed so far (Amelino--Camelia et al., 1997, 1998; Ellis, Mavromatos \& Nanopoulos, 2000; Ng \& van Dam, 1999) and eventually later criticized (Adler et al., 2000).

Most recently Lieu \& Hillman, 2002 (hereafter LH02) suggested that differential phase measurements of light propagated over a long distance, as implicitly made by interferometry of an extragalactic source, can place much tighter 
constraints on PSP.
They derived the effect on the random phase variation as depending upon the 
ratio of the photon wavelength $\lambda$ and the
Planck length $l_P$.

It is important to point out that the effects described by LH02 refers
to a model of PSP leading to random variations of the light phase, while
others exhibit a definite modification of radiation behavior for a
given wavelength (Jacobson, Liberati \& Mattingly, 2002;
Amelino--Camelia 2002). In other words any spacetime structure model
that yields a definite modification at a given wavelength is
unconstrained by the random phase approach. 

Furthermore, we are aware that LH02 has been the subject of even more
recent criticism (Ng, van Dam \& Christiansen, 2003) essentially based
upon the idea that such random perturbation of space--time should add
incoherently along the propagation path, leading to a square-root
dependency upon the distance between the source and the observer. 
We just here note that such assumption is at the basis of other PSPs
(Amelino Camelia, 1994) already ruled out by experimental verifications
(Ng \& van Dam, 1999) and that other theories does not incorporate such
dependencies too (Karolyhazy, 1966; Ng \& van Dam, 1994).

In this letter we argue that the use of diffraction, as an interferometry
effect by a telescope dish, can put stringent limits on the PSP with
random phase variations.

\section{Single aperture observations}

Following LH02 we assume that the error in the phase of a wavefront is just a different way of expressing the impossibility of measuring, by means of
light at wavelength $\lambda$, a distance $L$ with a precision $\Delta L$ such that:

\begin{equation}
\frac{\Delta L}{L} < a_0 \left( \frac{l_P}{\lambda} \right)^\alpha
\end{equation}

where the parameters $a_0$ and $\alpha$ characterize the theory being tested. For instance $\alpha =1/2$ corresponds to the random--walk approach (Amelino--Camelia, 2000) while $\alpha = 2/3$ corresponds to the holography principle (Wheeler, 
1982; Hawking, 1975) and $\alpha=1$ is the natural choice in a linearized theory. The coefficient $a_0$ can be reasonably expected to be of the order of the unity but, according to Amelino--Camelia (2001b) it can be {\it a few orders of magnitude} below 
unity. This gives an idea of the region of the parameter space described by $a_0$ and $\alpha$ where a meaningful search should be done.

\clearpage

\begin{figure}
\plotone{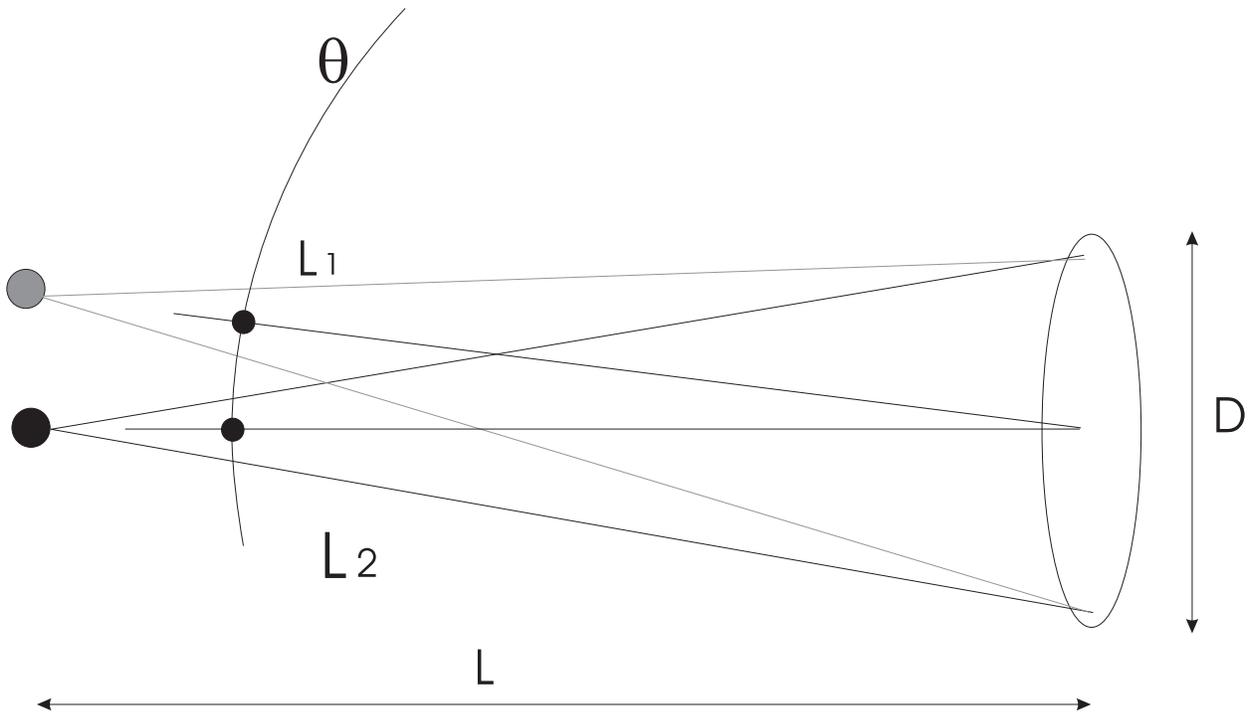}
\caption{The observation of a light source at a distance $L$ from the center of the telescope aperture.  The distances between the source and two extremity positions on the aperture are denoted by $L_1$ and $L_2$.  A variation in $L_1$ and
$L_2$ will result in an apparent displacement $\Delta \theta$ in the location of the source
.}
\end{figure}

\clearpage

Let us assume, in fact, that the measurement of {\it any} distance between a light source and a telescope of diameter $D$ is affected by the PSP described 
above, and translates into independent modification of the wavefront 
phase as determined from two distinct positions (this assumption, in the LH02 framework, corresponds to the, obviously verified, requirement of $D \gg l_P$). Let us consider the distances $L_1$ and $L_2$ as measured from a point source placed at a distance 
$L\approx L_1 \approx L_2$ from the two sides of a telescope of aperture $D$, see Fig.1. Any intrinsic  variation $\Delta L$ in the wavefront along the two lines of sight will translate into
 an apparent angular shift $\Delta \theta$ given by:

\begin{equation}
\Delta \theta \approx \frac{\Delta L}{D}
\end{equation}

where we did not co-add the (independent) uncertainties over the possible set of sightliness starting from any point of the telescope pupil (for instance, considering only $L_1$ and $L_2$ a $\sqrt{2}$ factor should be inserted into Eq.3) as this
 would not change the order of magnitude of the result.

It is important to emphasize that this result does not presume our knowledge of either $L_1$ or $L_2$ with the accuracy stated in Eq.(2). Actually the distance of any astronomical object is known with a much poorer accuracy than 
that required to 
test Eq.(2). The key point here is the {\it independence} of the accuracies on the measurements.   Specifically, the difference in the optical paths joining
various points of the telescope pupil and the (unresolved) source is
randomly modified by PSP.

If the PSP effects of Eq. (2) are present,  will this lead to a deterioration
of any interference pattern (e.g. the Airy rings of a filled aperture) 
seen at the diffraction limit ?  Such a consequence is
inevitable - to avoid it one must invoke the highly unlikely scenario of
{\it correlated} fluctuations in the optical paths  over all
points along the entire span
of the light footprint, which in general  has a size $\gg l_P$ (excepting
only an initial segment of paths, $\sim l_P \times L/D$ in length, which
for the  purpose of this work is an irrelevantly short 
(i.e. $\ll \lambda$) segment).

Thus it is reasonable to deduce that PSP leads to
an apparent angular broadening of a light source placed at a distance $L$, 
as seen from a telescope of diameter $D$, given by:

\begin{equation}
\Delta \theta = a_0 \frac{L}{D} \left( \frac{l_P}{\lambda} \right)^\alpha
\end{equation}

We compare such an angular broadening with the diffraction limit imposed by the telescope aperture by introducing a ratio $\eta$ defined as:

\begin{equation}
\eta = \frac{\Delta \theta}{\lambda/D} = a_0 \frac{L}{\lambda} \left( \frac{l_P}{\lambda} \right)^\alpha
\end{equation}

The meaning of $\eta$ is that it  directly influences
fringe visibility in the case of an interferometer, or the Strehl ratio $S$ 
of deterioration in point spread function
in the case of a telescope.  This is because 
one can write, following Sandler et al. 1994 and assuming that the broadening 
is equivalent to a blurring effect due to 
(e.g.) atmospheric disturbance, the following equation for $S$:

\begin{equation}
S = \exp\left( -\eta^2\right)
\end{equation}

It is reasonable to adopt $\eta = 1$ as rough criterion for any
experimental setup of this kind to secure a reliable test of PSP effects.  At
$\lambda =1 \mu$m, a representative wavelength for diffraction limited
optical telescopes (including the $D=2.4$m aperture of the
HST and $D=8\ldots 10$m class ground--based telescopes equipped with
Adaptive Optics facilities), this criterion requires the observation
of sources
at a minimum distance $L_{\rm min} \approx 6.2\times 10^{22}$ m 
$\approx 2.1$ Mpc, as already noted in LH02, to detect or to rule out the
case for $\alpha=1$, $a_0=1$.

\section{Astronomical benchmarks}

A celestial object that appears extended can either be because it is
genuinely so, or PSP causes a blurring of the image in the manner described
above.  To avoid confusion between the two possibilities, the best target 
choice is a Supernovae (SNe).  This is because for a distant SNe, its angular
size must remain considerably smaller than the  telescope
diffraction limit even if one assumes that the SNe shell has been expanding
steadily at the speed of light since the initial explosion.  Evidently, then,
our purpose of scrutinizing PSP will be fulfilled by an investigation of
an HST archival image of SN1994D, located at 
$L \approx 13.7$Mpc (Patat et al., 1996).

\clearpage

\begin{figure}
\plotone{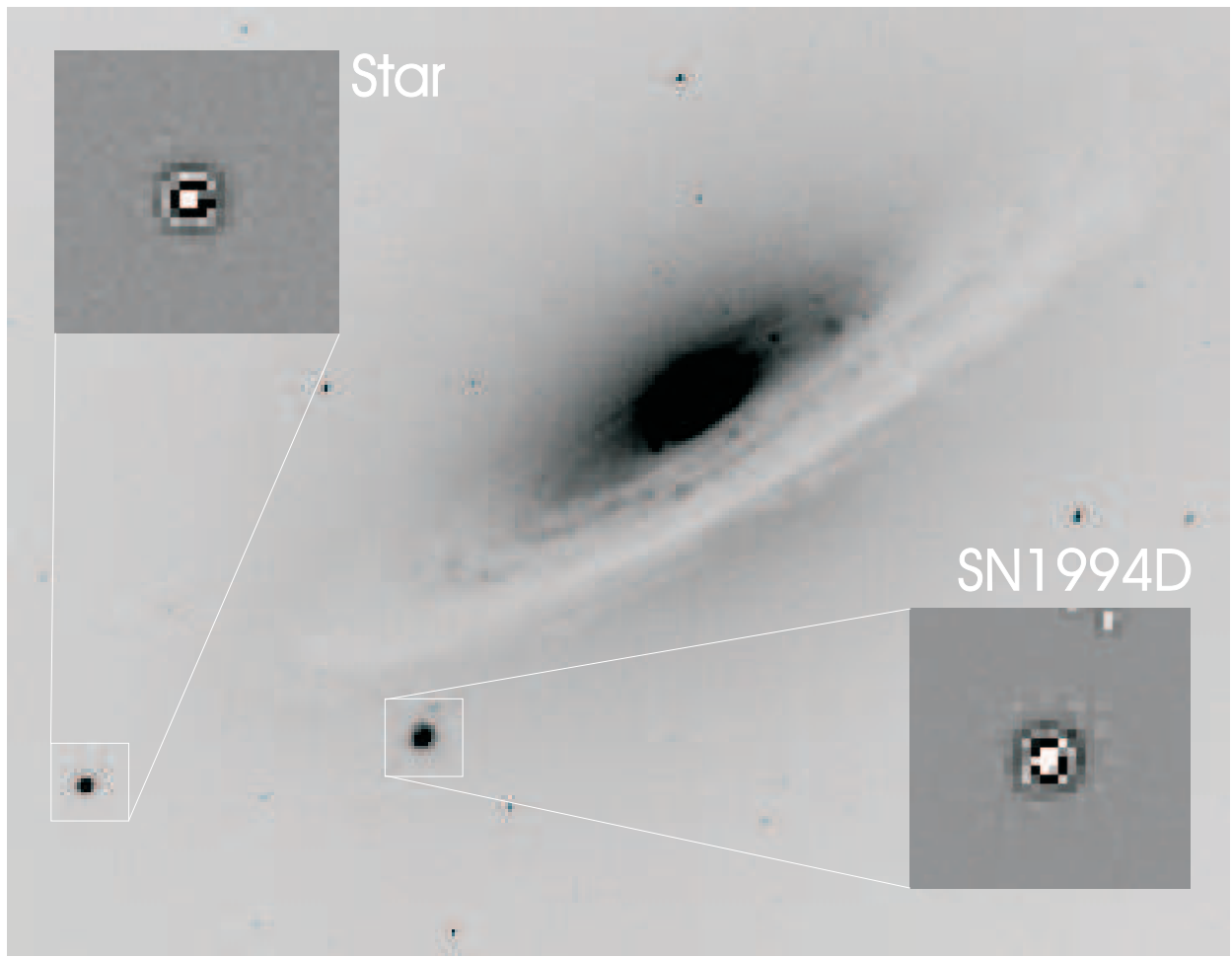}
\caption{SN1994D as taken with HST. In the boxes a Galactic star and the SN are shown with the high spatial frequencies content of both original images enhanced in the same way by subtracting a smoothed version (with a $3\times 3$ boxcar) of the same area. The Airy disks are clearly seen in both images, in spite of a small pixel size (equivalent to 0.046 arcsec) giving a poor sampling of the diffraction limit.}
\end{figure}

\clearpage

By comparing
the HST--collected frame of SN1994D with that of
a foreground Galactic star in the same field (see Fig.2) we see that both objects
exhibit no deterioration of their Airy interference patterns.  This
constrains the Strehl deterioration parameter to $S > 0.2$ (else
the first Airy rings will become invisible), and
hence (by Eq. 6) places a lower limit for $\eta$ at 
$\eta > 1.3$. 

A separate investigation concerns
the Hubble Deep Field high--$z$ images.
Spectroscopic follow--ups have shown that objects 
as distant as $z=5.34$, corresponding to $L \approx 7.7$Gpc  are as small as 0.12arcsec (Spinrad et al., 1998).
The distance adopted here is the comoving radial distance, as it is the summation over the journey of the photon of the length experienced by a comoving observer, where we assume the PSP exhibits in the same way.
The exact value for $L$ depends on the cosmological model used; here we have chosen $H_0 = 72$km/s/Mpc and $(\Omega_M, \Omega_\Lambda) = (0.3,0.7)$ 
as given in Krauss (2002),

Since $\eta = 1.7$ is the ratio of this observed size to
the diffraction limit capability of a $D=2.4$m telescope at the relevant
wavelength of $\lambda =814$nm, one can clinch PSP even further than before.
We note in passing that, in reality, such a ratio for $\eta$ understates the
case against PSP, because while propagating
from the source to us a photon initially had shorter wavelength, 
so that the quantity $\Delta L$ of Eq.(2),
hence $\Delta \theta$ of Eq. (3), was larger in the past.

\clearpage

\begin{figure}
\plotone{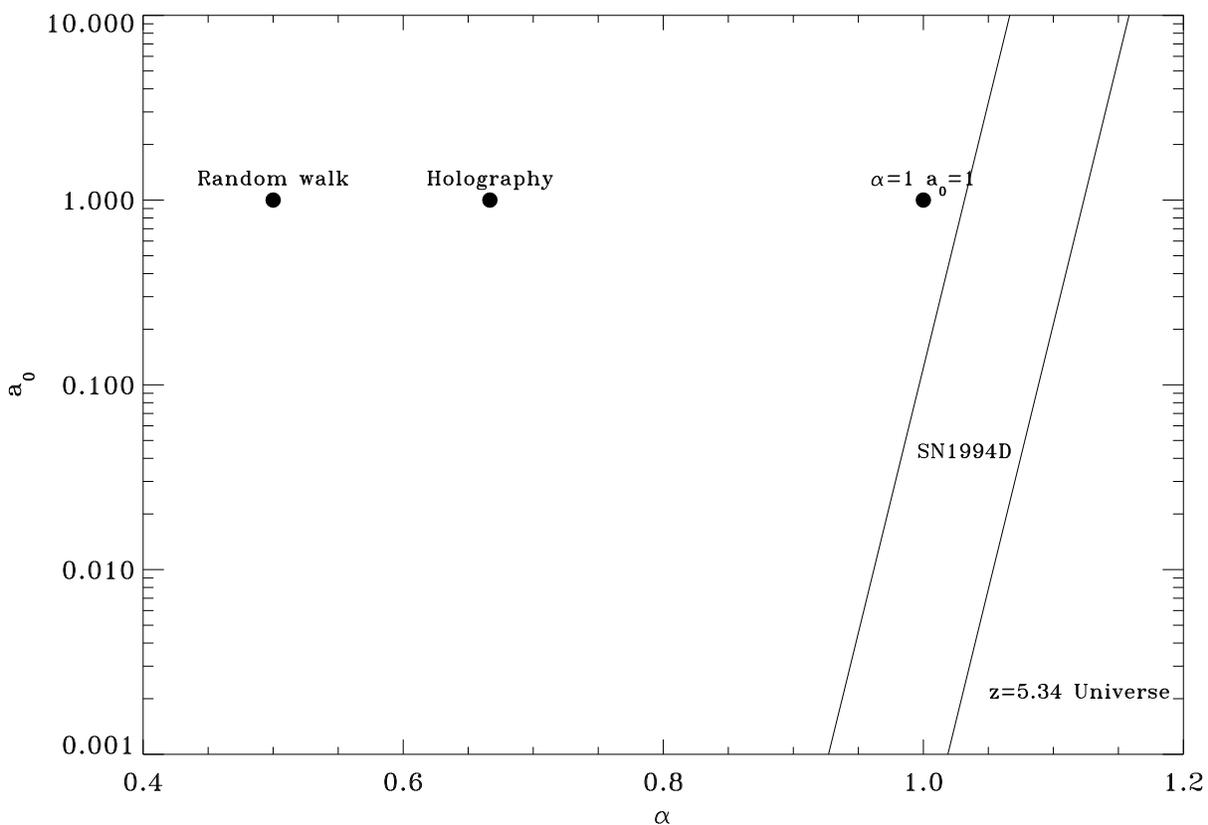}
\caption{A portion of the parameter space $a_0 - \alpha$. The constraints imposed by the observations discussed in the text allow only the regions on the right side of the two oblique lines. Three points representing different PSP parameter choices are also shown, where we assumed the coupling coefficient $a_0 =1$.}
\end{figure}

\clearpage

\section{Discussion}

In Fig. 3 implications of the two observations
being analyzed thus far are plotted in $a_0$--$\alpha$ space.
We can see that a linear, first--order PSP characterized by $\alpha =1, a_0=1$
is consistently excluded, as are the other cited phenomenologies with smaller $\alpha$, for all reasonable
values of $a_0$. In particular, when $\alpha =1$, the upper limit on the angular size of
high--$z$ objects requires that $a_0 < 3\times 10^{-4}$. 

The two benchmarks presented in the previous section were
established with the most powerful  instruments currently available (HST for measurement of the angular size and Keck for determination of the cosmological distance).  It should be realized that, in general, the existence of
PSP with $\alpha \sim a_0\sim 1$ would render the universe unobservable at any appreciable redshift, due to the significant blurring of the images of point sources. This may be regarded as a form of Olber's paradox.  In the case of
the far universe, where observations require special technique, additional
benchmarks could be envisaged.

Quantitatively, the limits given above for the exclusion of first order
PSP understate the case, because (a) the errors were estimated conservatively -
they would have assumed larger values had we propagated them at every step;
(b) the wavelength of radiation from a distant source is shorter towards
the source, meaning that our upper limit on $a_0$ should in reality be
even smaller.  Thus, in the same context as that of LH02, the possibility of
$\alpha=1$ may be ruled out with confidence.

Our conclusions may be compared and contrasted with other recent works,
notably those of Jacobson, Liberati \& Mattingly (2002) and
Amelino--Camelia (2002).  The former 
used X-ray observations of the Crab nebula to argue against PSP, its
validity depends on the assumptions made about the physical
processes in the Crab.   The latter, however, proposed the {\it existence} of 
PSP effects as the reason why gamma-rays from a distant quasar survive
their journey through the intergalactic medium to reach us.  We note
here also, that alternative interpretations are entirely possible.

In the framework of the assumptions made in LH02,
PSP effects are excluded by the observations described in this Letter.
Perhaps there exist some {\it ad hoc} explanations as to why first order
PSP cannot
be manifested 
as perturbation of a light pencil.   As regards whether the present 
findings imply that the notion of structural space time at the Planck scales (a sort of {\it aether} embedded in the continuum where familiar physics holds) is untenable, or
whether a subtle mechanism is at play to render such structures evasive,
these questions are outside the scope of our Letter.

\section*{Acknowledgements}
Thanks are due for the useful hints and discussions to Richard Lieu
and Lauro Moscardini. We also thank the unknown referee for his
helpful comments.

\section*{References}

\begin{itemize}
\item Adler R.J., Nemenman I.M, Overduin J.M., Santiago D.I. 2000, Phys.Lett. B477 24-428
\item Adler R.J., Santiago D.I. 1999, Mod.Phys.Lett. A14, 1371
\item Amelino--Camelia G., Ellis J., Mavromatos N.E., Nanopoulos D.V.1997, Int.J.Mod.Phys. A12, 607-624
\item Amelino--Camelia G. 1994, Mod. Phys. Lett. A9, 3415 
\item Amelino--Camelia G., Ellis J., Mavromatos N.E., Nanopoulos D.V., Sarkar S. 1998, Nature 393, 763
\item Amelino--Camelia G. 2000, Towards quantum gravity, Proc. of the XXXV International    Winter school on Theor. Phys., Polanica, Poland, Ed. Jerzy Kowalski--Glikman.       Lecture Notes in Physics 541, 1 Berlin: Springer--Verlag
\item Amelino--Camelia G. 2001a  Int.J.Mod.Phys. D10 (2001) 1-8
\item Amelino--Camelia G. 2001b Nature 410, 1065
\item Amelino--Camelia G. 2002 Proceedings of the Ninth Marcel Grossmann Meeting on General Relativity,
      edited by V.G. Gurzadyan,  R.T. Jantzen and R. Ruffini,
      World Scientific, Singapore, 1485 (also gr-qc/0106005)
\item Ellis J., Mavromatos N.E., Nanopoulos D.V. 2000, Gen. Rel. Grav. 32, 127
\item Garay L.J. 1998,  Phys.Rev.Lett. 80, 2508
\item Jacobson T., Liberati S., Mattingly D. 2002 astro-ph/0212190
\item Hawking S. 1975, Comm. Math. Phys. 43, 199
\item Karolyhazy K. 1966, Il Nuovo Cimento A42, 390
\item Kempf A. 1999, Rept. Math. Phys. 43, 171
\item Krauss L.M. 2002 in Proc. ESO--CERN--SAS Symp. on Astronomy, Cosmolgy and Fundamental Physics, March 2002, also astro-ph/0301012
\item Lieu R., Hillman L.W. 2002 astro-ph/0211402 ApJL in press (LH02 in the text)
\item Ng Y.J., van Dam H. 1994, Mod. Phys. Lett. A9, 335 
\item Ng Y.J., van Dam H. 1999, Found. Phys. 30, 795
\item Ng Y.J., van Dam H., Christiansen W.A. 2003, astro-ph/0302372
\item Patat F., Benetti S., Cappellaro E., Danziger I.J., della Valle M., Mazzali P.A., Turatto M. 1996, MNRAS 278, 111
\item Rovelli C. 1998, Living Rev. Rel. 1, 1
\item Salecker H., Wigner E.P. 1958, Physical Rev. 109, 571
\item Sandler D.G., Stahl S., Angel J.R.P., Lloyd--Hart M., McCarty D. 1994, JOSA--A 11, 925
\item Spinrad H., Stern D., Bunker A., Dey A., Lanzetta K., Yahil A., Pascarelle S., Fernandez--Soto A. 1998, AJ 116, 2617
\item Wheeler J. 1982, Int. J. Theor. Phys. 21, 557
\item Wigner E.P. 1957, Rev. Mod. Phys. 29, 255

\end{itemize}

\end{document}